
\documentstyle{amsppt}

\NoBlackBoxes
\magnification=\magstep1
\hsize=6.5truein
\vsize=8.5truein
\document
\baselineskip=.15truein

\topmatter
\title   The  $W_{1+\infty}(gl_s)$--symmetries of the $S$--component KP
hierarchy
\endtitle
\author
Johan van de Leur
\endauthor
\thanks
The research of Johan van de Leur is
financially supported by the ``Stichting Fundamenteel Onderzoek der
Materie (F.O.M.)''. E-mail: vdleur\@math.utwente.nl
\endthanks
\abstract
Adler, Shiota and van Moerbeke obtained for the KP and Toda lattice
hierarchies
a formula which translates  the action of the
vertex operator on tau--functions to an action of a vertex operator
of pseudo-differential
operators on  wave functions. This  relates  the additional
symmetries of the KP and Toda lattice hierarchy
to the $W_{1+\infty}$--, respectively $W_{1+\infty}\times
W_{1+\infty}$--algebra
symmeties.  In this paper we generalize the
results to the $s$--component KP hierarchy. The vertex operators
generate the algebra $W_{1+\infty}(gl_s)$, the matrix version of
$W_{1+\infty}$. Since the Toda lattice hierarchy is equivalent to the
$2$--component KP hierarchy, the results of this paper uncover in that
particular
case a much richer structure than the one obtained by Adler, Shiota
and van Moerbeke.

\endabstract
\endtopmatter

\vskip 10pt

\subheading{\S 0. Introduction}
\vskip 10pt
\noindent The KP hierarchy is the set of deformation equations
$${\partial L\over\partial t_k}=[(L^n)_+,L],$$
for the first order pseudo-differential operator
$$L\equiv
L(x,t)=\partial+u_1(x,t)\partial^{-1}+u_2(x,t)\partial^{-2}+\cdots,$$
here $\partial={\partial\over\partial x}$ and $t=(t_1,t_2,\ldots)$.
It is well--known that $L$ dresses as $L=P\partial P^{-1}$ with
$$\aligned
P\equiv
P(\tau,x,t)&=1+a_1(x,t)\partial^{-1}+a_2(x,t)\partial^{-2}+\cdots\\
&={\tau(x,t-[\partial^{-1}])\over \tau(x,t)},
\endaligned
$$
where $\tau $ is the famous $\tau$--function, introduced by the Kyoto
group [DJKM1-3] and $[z]=(z,{z^2\over 2},{z^3\over 3},\ldots)$.

The wave or Baker--Akhiezer function
$$\Psi\equiv \Psi(\tau,x,t,z)= W(\tau,x,t,\partial)e^{zx},$$
where
$$W\equiv W(\tau,x,t,z)=P(\tau,x,t)e^{\xi(t)}\qquad
\text{with}\quad\xi(t)=\sum_{k=1}^\infty
t_k\partial^k$$
is an eigenfunction of L, viz.,
$$L\Psi=z\Psi\qquad\text{and}\quad {\partial\Psi\over\partial
t_k}=(L^k)_+\Psi.$$
{}From this point of view, the introduction by Orlov and  Schulman [OS]
of another pseudo-differential operator $M\equiv M(x,t)=WxW^{-1}$ which action
on $\Psi$ amounts to
$$M\Psi={\partial\Psi\over\partial z}$$
is rather natural.

Recently, Adler, Shiota and van Moerbeke [ASV1], [ASV2] proved a
conjecture of Orlov and Schulman, viz.,  that there
exists a relation between $(M^\ell L^{k+\ell})_-$ acting on $\Psi$
and the generators $W_k^{(\ell+1)}\sim -t^{k+\ell}({\partial\over\partial
t})^{\ell}$
of the $W_{1+\infty}$--algebra  acting on the $\tau$--function. More
explicitly, let
$$Y(\tau,y,w)=\sum_{\ell=0}^\infty {(y-w)^\ell\over \ell !}\sum_{k\in\Bbb
Z}M^\ell L^{k+\ell} w^{-k-\ell-1}$$
be the generating series of the $M^\ell L^{k+\ell}$ and let
$$\aligned
W(y,w)&=\sum_{\ell=0}^\infty {(y-w)^\ell\over \ell !}\sum_{k\in\Bbb
Z}W^{(\ell+1)}_k w^{-k-\ell-1}\\
&={1\over y-w}e^{x(y-w)+\sum_{k=1}^\infty t_k(y^k-w^k)}
e^{-\sum_{k=1}^\infty {\partial\over\partial t_k}{y^{-k}-w^{-k}\over
k}}
\endaligned
\tag{0.1}
$$
be the vertex operator of the KP hierarchy, then one has the following
formula [ASV1]:
$$-Y(\tau,y,w)_-\Psi(\tau,x,t,z)=\Psi(\tau,x,t,z)(e^{-\sum_{k=1}^\infty
{\partial\over\partial t_k}{z^{-k}\over k}}-1)\left ( {W(y,w)\tau(x,t)\over
\tau(x,t)}\right ).\tag{0.2}$$
Dickey gave another proof of this formula [D]. The ``geometric
interpretation'' of this Adler--Shiota--van Moerbeke  formula is as follows.
The transformation $e^{\lambda W(y,w)}=1+\lambda W(y,w)$ is a symmetrie
transformation
or a kind of auto-B\"acklund
transformation of the KP hierarchy. If one rewrites (0.2) as
$$
-\lambda Y(\tau,y,w)_-\Psi(\tau,x,t,z)=\Psi(\tau,x,t,z)(e^{-\sum_{k=1}^\infty
{\partial\over\partial t_k}{z^{-k}\over k}}-1)\left ({e^{\lambda
W(y,w)}\tau(x,t)\over
\tau(x,t)}\right ),\tag{0.3}$$
then one easily sees that  (0.2) is in fact a formula that relates this
B\"acklund transformation to the so--called additional
symmetries of the KP hierarchy. To be more precise, let $\sigma$ be
the new solution of the KP hierarchy which one obtains from $\tau$ by
this B\"acklund transformation, i.e., $\sigma =e^{\lambda
W(y,w)}\tau$, then
$$\aligned
\Psi(\sigma,x,t,z)&=\sigma(x,t)^{-1}e^{-\sum_{k=1}^\infty
{\partial\over\partial t_k}{z^{-k}\over
k}}\sigma(x,t)e^{\xi(t)}e^{zx}\\
&={\tau(x,t)\over\sigma(x,t)}e^{-\sum_{k=1}^\infty
{\partial\over\partial t_k}{z^{-k}\over k}}\left ({e^{\lambda
W(y,w)}\tau(x,t)\over
\tau(x,t)}\right )\Psi(\tau,x,t,z)\\
&=\Psi(\tau,x,t,z)+{\tau(x,t)\over\sigma(x,t)}(e^{-\sum_{k=1}^\infty
{\partial\over\partial t_k}{z^{-k}\over k}}-1)\left ({e^{\lambda
W(y,w)}\tau(x,t)\over
\tau(x,t)}\right )\Psi(\tau,x,t,z).
\endaligned$$
Now using (0.3) one obtains
$$\Psi(\sigma,x,t,z)=\left (
1-\lambda{\tau(x,t)\over\sigma(x,t)}Y(\tau,y,w)_-\right
)\Psi(\tau,x,t,z).$$
Hence $Y(\tau,y,w)_-$ produces, as a consequence of formula (0.2), the
B\"acklund transformation of the wave function corresponding to $\tau$.

Adler, Shiota and van Moerbeke also treated in [ASV2] the Toda lattice
hierarchy of Ueno and Takasaki [UT] and showed that an analogous
formula also holds. In their treatment they considered two vertex
operators, each depending on a different time flow
$t^{(j)}=(t^{(j)}_1, t^{(j)}_2,\dots)$, $j=1,2$,
of a form similar to that of (0.1). Hence The $W_{1+\infty}$--algebra
of the KP hierarchy is replaced by two copies of this algebra.

Using $2\times 2$--matrix pseudo--differential operators instead of
infinite shift operators, one can show that the Toda lattice hierarchy
is equivalent to the 2--component KP hierarchy as treated by Kac and
the author in [KV]. In that case there are however more vertex
operators than only the ones of the form (0.1), viz., one also has
$$W^{(ab)}(y,w)={C^{(ab)}(y,w)\over(y-w)^{\delta_{ab}}}e^{x(y-w)+\sum_{k=1}^\infty (t^{(a)}_ky^k-t^{(b)}_kw^k)}
e^{-\sum_{k=1}^\infty{1\over k} ({\partial\over\partial t^{(a)}_k}y^{-k}-
{\partial\over\partial t^{(b)}_k}w^{-k})},$$
here $C^{(ab)}(y,w)$ are operators that act on the twisted group
algebra of the root lattice of $sl_2$. A natural question now is: Are
there also matrix pseudo--differential operators such that a formula
as (0.2) hold for these $W^{(ab)}(y,w)$?  In this paper we show that
a similar result holds, not only for the 2--component KP hierarchy, but
in general for the $s$--component KP hierarchy. One finds that the
natural generalization of $W_{1+\infty}$ is not $(W_{1+\infty})^s$ but
$W_{1+\infty}(gl_s)$, the central extension of the algebra of
differential operators on $(\Bbb C[t,t^{-1}])^s$. Hence one can
conclude that the results of [ASV2] for the Toda lattice hierarchy are
not complete, but that the structure is richer.

It is a pleasure to thank Gerard Helminck, Peter van den Heuvel
and Takahiro Shiota,  for usefull discussions.

\vskip 10pt
\subheading{\S 1.  $a_{\infty}$ and the KP hierarchy in
the fermionic picture [KV]}

\vskip 10pt
\noindent
Consider the infinite dimensional complex  Lie algebra
$a_{\infty} := \overline{gl_{\infty}}\bigoplus {\Bbb C}c$, where
$$\overline{gl_{\infty}} = \{ a = (a_{ij})_{i,j \in {\Bbb Z}+\frac{1}{2}}
|\  a_{ij}=0\  \text{if}\   |i-j|>>0\},$$
with Lie bracket defined by
$$[a+\alpha c,b+\beta c]=ab-ba+\mu (a,b)c , \tag{1.1}$$
for $a,b\in \overline{gl_{\infty}}$ and $\alpha ,\beta\in{\Bbb C}$.
Here $\mu$ is the following 2--cocycle:
$$\mu (E_{ij},E_{kl})=\delta_{il}\delta_{jk}(\theta(i)-\theta(j)),
\tag{1.2}$$
with $E_{ij}$ the matrix that has a $1$ on the $(i,j)$-th entry and zeros
elsewhere and $\theta :{\Bbb R}\to {\Bbb C}$ the function  defined by
$$\theta(i):=\cases 0 & \text{if}\ i>0,\\
                    1 & \text{if}\ i\le 0.\endcases \tag{1.3}$$
Let ${\Bbb C}^{\infty} = \bigoplus_{j \in {\Bbb Z}+\frac{1}{2}} {\Bbb
C} v_{j}$ be the infinite dimensional complex vector space with fixed
basis $\{ v_{j}\}_{j \in {\Bbb Z}+\frac{1}{2}}$. The Lie algebra $a_{\infty}$
acts linearly on
${\Bbb C}^{\infty}$ via the usual formula:
$$E_{ij} (v_{k}) = \delta_{jk} v_{i}.$$
We introduce,
following [KP2], the corresponding semi-infinite wedge space $F =
\Lambda^{\frac{1}{2}\infty} {\Bbb C}^{\infty}$, this is the vector space
with a basis consisting of all semi-infinite monomials of the form
$v_{i_{1}} \wedge v_{i_{2}} \wedge v_{i_{3}} \ldots$, where $i_{1} >
i_{2} > i_{3} > \ldots$ and $i_{\ell +1} = i_{\ell} -1$ for $\ell >>
0$.
In order to describe representations of the Lie algebra on this space,
 we find it convenient to define  wedging and contracting
operators $\psi^{-}_{j}$ and $\psi^{+}_{j}\ \ (j \in {\Bbb Z} +
\frac{1}{2})$ on $F$ by
$$\align
&\psi^{-}_{j} (v_{i_{1}} \wedge v_{i_{2}} \wedge \cdots ) = \cases 0
& \text{if}\ j = -i_{s} \text{ for some}\ s \\
(-1)^{s} v_{i_{1}} \wedge v_{i_{2}} \cdots \wedge v_{i_{s}} \wedge
v_{-j} \wedge v_{i_{s+1}} \wedge \cdots &\text{if}\ i_{s} > -j >
i_{s+1}\endcases \\
&\psi^{+}_{j} (v_{i_{1}} \wedge v_{i_{2}} \wedge \cdots ) = \cases 0
&\text{if}\ j \neq i_{s}\ \text{for all}\ s \\
(-1)^{s+1} v_{i_{1}} \wedge v_{i_{2}} \wedge \cdots \wedge
v_{i_{s-1}} \wedge v_{i_{s+1}} \wedge \cdots &\text{if}\ j = i_{s}.
\endcases
\endalign
$$
These  wedging and contracting operators satisfy the following relations
$(i,j \in {\Bbb Z}+\frac{1}{2}, \lambda ,\mu = +,-)$:
$$\psi^{\lambda}_{i} \psi^{\mu}_{j} + \psi^{\mu}_{j}
\psi^{\lambda}_{i} = \delta_{\lambda ,-\mu} \delta_{i,-j}, \tag{1.4}$$
hence they generate a Clifford algebra, which we denote by ${\Cal C}\ell$.

Introduce the following elements of $F$ $(m \in {\Bbb Z})$:
$$|m\rangle = v_{m-\frac{1}{2} } \wedge v_{m-\frac{3}{2} } \wedge
v_{m-\frac{5}{2} } \wedge \cdots .$$
It is clear that $F$ is an irreducible ${\Cal C}\ell$-module such that
$\psi^{\pm}_{j} |0\rangle = 0 \ \text{for}\ j > 0 . $
Define  a representation $\hat r$ of $a_{\infty}$ on $F$ by
$$ \hat r(E_{ij})=:\psi^-_{-i}\psi^+_j:,\quad \hat r(c)=I,\tag{1.5}
$$
where $:\ :$ stands for the {\it normal ordered product} defined in
the usual way $(\lambda ,\mu = +$ or $-$):
$$:\psi^{\lambda (i)}_{k} \psi^{\mu (j)}_{\ell}: = \cases \psi^{
\lambda (i)}_{k}
\psi^{\mu (j)}_{\ell}\ &\text{if}\ \ell \ge k \\
-\psi^{\mu (j)}_{\ell} \psi^{\lambda (i)}_{k} &\text{if}\ \ell <
k.\endcases \tag{1.6}$$

Define the {\it charge decomposition}
$$F = \bigoplus_{m \in {\Bbb Z}} F^{(m)} \tag{1.7}$$
by letting

$$\text{charge}(|0\rangle ) = 0\ \text{and charge} (\psi^{\pm}_{j}) =
\pm 1. \tag{1.8}$$
It  is easy to see that each $F^{(m)}$ is an  irreducible $a_{\infty}$--highest
weight module with
highest weight vector $|m\rangle$.

We are now able to define the {\it KP hierarchy in the fermionic
picture}, it is the equation
$$\sum_{k \in {\Bbb Z}+\frac{1}{2}} \psi^{+}_{k} \tau \otimes \psi^{-}_{-k}
\tau = 0, \tag{1.9}$$
for
$\tau\in F^{(0)}$. One can prove (see e. g. [KP2] or [KR]) that
this equation characterizes  the group orbit of the vacuum vector
$|0\rangle$  for the  the group $GL_{\infty}$.
Since the group does not play an important role in this paper,
we will not introduce it here.
\vskip 10pt

\subheading{\S 2. $W_{1+\infty}(gl_s)$ as subalgebra of $a_{\infty}$}

\vskip 10pt
\noindent Let $e_i$, $1\le i\le s$ be a basis of ${\Bbb C}^s$.
By identifying $({\Bbb C}[t,t^{-1}])^s$ with ${\Bbb C}^{\infty}$,
we can embed   $W_{1+\infty}(gl_s)$ into  $a_{\infty}$. This,
however, can be done in many different ways, the simplest
one is the following.
We put ($1\le a \le s,\ j\in {\Bbb Z} + \frac{1}{2})$:
$$\align
v^{(a)}_{j}&=v_{sj+\frac{1}{2}(s-2a+1)} =t^{-j-{1\over 2}}e_a,\\
\psi^{\pm (a)}_{j }&=\psi^{\pm }_{sj\pm\frac{1}{2}(s-2a+1)}.
\tag{2.1}
\endalign$$
Notice that with this relabeling we have:
$\psi^{\pm (a)}_{k}|0\rangle = 0\ \text{for}\ k > 0.$
We introduce the  generating series of the fermions, the so--called
fermionic fields $(z \in {\Bbb C}^{\times})$:
$$\psi^{\pm (a)}(z) \overset{\text{def}}\to{=} \sum_{k \in {\Bbb
Z}+\frac{1}{2}} \psi^{\pm
(a)}_{k} z^{-k-\frac{1}{2}}.\tag{2.2}$$
We also rewrite the $E_{jk}$'s:
$$E^{(ab)}_{j,k}=E_{sj+\frac{1}{2}(s-2a+1),sk+\frac{1}{2}(s-2b+1)},
\tag{2.3}$$
then  $\hat r(E^{(ab)}_{jk})=:\psi^{-(a)}_{-j}\psi^{+(b)}_k:$.

We can associate to $({\Bbb C}[t,t^{-1}])^s$  the Lie algebra
of differential operators on this space,
it has as basis the operators:
$$-t^{k+\ell}({\partial\over\partial t})^{\ell}e_{ij},
\quad\text{for}\ k\in {\Bbb Z},\ \ell\in{\Bbb Z}_+,\ 1\le i,j\le s.$$
We will denote this Lie algebra by $D(gl_s)$. We can embed this algebra via
(2.1)
into $\overline{gl_{\infty}}$
and also into $a_{\infty}$, one finds
$$-t^{k+\ell}({\partial\over\partial t})^{\ell}e_{ij}
\mapsto \sum_{m\in{\Bbb
Z}}-m(m-1)\cdots(m-\ell+1)E^{(ij)}_{-m-k-\frac{1}{2},-m-\frac{1}{2}}.
\tag{2.4}
$$
It is straightforward, but rather tedious, to calculate the corresponding
2--cocycle, the result is as follows (see also [KP1], [Ra] and [KRa]).
Let $f(t), g(t)\in {\Bbb C}[t,t^{-1}]$ and $a,b\in gl_s$ then
$$\mu(f(t)({\partial\over\partial t})^{\ell}a,g(t)({\partial\over\partial t})^m
b)
={{\ell !m!}\over(\ell+m+1)!}\text{Res}_{t=0}dt\
f^{(m+1)}(t)g^{(\ell)}(t)\text{trace}(ab).
$$
Hence in this way we get a central extension $W_{1+\infty}(gl_s)=D(gl_s)\oplus
{\Bbb C}c$ of $D(gl_s)$
with Lie bracket
$$\align
&[f(t)({\partial\over\partial t})^{\ell}a+\alpha c,g(t)({\partial\over\partial
t})^m b+\beta c]=\\
&f(t)({\partial\over\partial t})^{\ell}g(t)({\partial\over\partial
t})^m
 ab-
g(t)({\partial\over\partial t})^m f(t)({\partial\over\partial t})^{\ell}ba
+\mu(f(t)({\partial\over\partial t})^{\ell}a,g(t)({\partial\over\partial t})^m
b)c.
\tag{2.5}\endalign
$$
Since we have the representation $\hat r$ of $a_{\infty}$, we find that
$$\hat r (-t^{k+\ell}({\partial\over\partial t})^{\ell}e_{ab})=
\sum_{m\in{\Bbb Z}}m(m-1)\cdots (m-\ell+1):\psi^{+(a)}_{-m-\frac{1}{2}}
\psi^{-(b)}_{m+k+\frac{1}{2}}:.
$$
In terms of the fermionic fields (2.2), we find
$$\sum_{k\in{\Bbb Z}}\hat r (-t^{k+\ell}({\partial\over\partial
t})^{\ell}e_{ab})z^{-k-\ell-1}=
:{{\partial^{\ell}\psi^{+(a)}(z)}\over \partial z^{\ell}}
\psi^{-(b)}(z):. \tag{2.6}
$$
Now define
$$\aligned
W^{(ab,\ell+1)}_k:&=\hat r (-t^{k+\ell}({\partial\over\partial
t})^{\ell}e_{ab}),\\
W^{(ab,\ell+1)}(z)&=\sum_{k\in{\Bbb Z}}W^{(ab,\ell+1)}_k z^{-k-\ell-1},
\endaligned
\tag{2.7}$$
then
$$\aligned W^{(ab)}(y,z):&=:\psi^{+(a)}(y)
\psi^{-(b)}(z):\\ &=\sum_{\ell =0}^\infty {(y-z)^\ell \over \ell !}
W^{(ab,\ell+1)}(z)\\
&= \sum_{\ell =0}^\infty {(y-z)^\ell \over \ell !}\sum_{k\in\Bbb Z}
W^{(ab,\ell+1)}_kz^{-k-\ell-1}.\endaligned
\tag{2.8}$$

\subheading{\S 3. The $s$--component boson fermion correspondence}

\vskip 10pt
\noindent
Using a bosonization one can rewrite (1.9) as a system
of partial
differential equations and express the basis elements of
$W_{1+\infty}(gl_s)$ in terms of vertex operators.
 We begin by introduce bosonic fields $(1 \leq i \leq s)$:
$$\alpha^{(i)}(z) \equiv \sum_{k \in {\Bbb Z}} \alpha^{(i)}_{k} z^{-k-1}
\overset{def}\to{=} :\psi^{-(i)}(z) \psi^{+(i)}(z):, \tag{3.1}$$
Since $\alpha^{(i)}(z)=W^{(ii,1)}(z)$, one easily checks  that the operators
$\alpha^{(i)}_{k}$ satisfy
the canonical commutation relation of the associative
oscillator algebra,  which we
denote by ${\frak a}$:
$$[\alpha^{(i)}_{k},\alpha^{(j)}_{\ell}] =
k\delta_{ij}\delta_{k,-\ell},\tag{3.2}$$
and one has
$$\alpha^{(i)}_{k}|m\rangle = 0 \ \text{for}\ k > 0.\tag{3.3}$$

It is easy to see that restricted to $\hat{gl}_{s}$, which is the subalgebra
generated
by the elements $W^{(ij,1)}_k$,
$F^{(0)}$ is its basic highest weight representation (see [K, Chapter
12]).

We will now
describe the $s$-component boson-fermion
correspondence (see [KV]).    Let $L$ be a lattice with a basis
$\delta_{1},\ldots
,\delta_{s}$ over ${\Bbb Z}$ and the symmetric bilinear form
$(\delta_{i}|\delta_{j}) = \delta_{ij}$, where $\delta_{ij}$ is the
Kronecker symbol.  Let
$$\varepsilon_{ij} = \cases -1 &\text{if $i > j$} \\
1 &\text{if $i \leq j$.} \endcases \tag{3.4}$$
Define a bimultiplicative function $\varepsilon :\ L \times L @>>> \{
\pm 1 \}$ by letting
$$\varepsilon (\delta_{i}, \delta_{j}) = \varepsilon_{ij}.
\tag{3.4}$$
Let $\delta = \delta_{1} + \ldots + \delta_{s},\  Q= \{ \gamma \in
L|\ (\delta | \gamma ) = 0\}$, $\Delta = \{ \alpha_{ij} :=
\delta_{i}-\delta_{j}| i,j = 1,\ldots ,s,\ i \neq j \}$.  Of course
$Q$ is the root lattice of $sl_{s}({\Bbb C})$, the set $\Delta$
being the root system.

Consider the vector space ${\Bbb C}[L]$ with basis $e^{\gamma}$,\
$\gamma \in L$, and the following twisted group algebra product:
$$e^{\alpha}e^{\beta} = \varepsilon (\alpha ,\beta)e^{\alpha +
\beta}. \tag{3.6}$$ Let ${\Bbb C}[t]$ be the space of polynomials in
indeterminates $t = \{ t^{(i)}_{k}\},\ k = 1,2,\ldots ,\ i =
1,2,\ldots ,s$ and denote by $B = {\Bbb C}[t] \otimes_{\Bbb C} {\Bbb C}[L]$ be
the tensor
product of these algebras.  Then the $s$-component boson-fermion
correspondence is the vector space isomorphism
$$\sigma :F @>\sim >> B, \tag{3.7}$$
given by $\sigma (|0\rangle )=1$ and
$$\sigma\psi^{\pm (a)}(z) \sigma^{-1}=e^{\pm \delta_a}z^{\pm \delta_a}
\exp(\pm\sum_{k=1}^{\infty}t_k^{(a)}z^k)\exp(\mp\sum_{k=1}^{\infty}
{\partial\over \partial t_k^{(a)}}{z^{-k}\over k}),\tag {3.8}$$
where
$$\delta_a (p(t) \otimes e^{\gamma}) =
(\delta_{a}|\gamma ) p(t) \otimes e^{\gamma}.\tag{3.9}$$
The transported charge then is as follows:
$$
\text{charge}(p(t)\otimes e^{\gamma}) = (\delta |\gamma).
$$
We denote the transported charge decomposition by
$B = \bigoplus_{m \in {\Bbb Z}} B^{(m)}$, then
the transported action of the operators $\alpha^{(i)}_{m}$ is given by
$$\cases
\sigma \alpha^{(j)}_{-m}\sigma^{-1}(p(t) \otimes e^{\gamma}) =
mt^{(j)}_{m}p(t)\otimes e^{\gamma},\ \text{if}\ m > 0, &\  \\
\sigma \alpha^{(j)}_{m} \sigma^{-1}(p(t) \otimes e^{\gamma}) =
 \frac{\partial
p(t)}{\partial t_{m}} \otimes e^{\gamma},\ \text{if}\ m > 0, &\  \\
\sigma \alpha^{(j)}_{0} \sigma^{-1} (p(t) \otimes e^{\gamma}) =
(\delta_{j}|\gamma ) p(t) \otimes e^{\gamma}
. & \
\endcases \tag{3.10}
$$

If one substitutes (3.8) into (2.8), one obtains  the following
vertex operator expression for the generating
series of the fields $W^{(ab,\ell+1)}(z)$:
$$\aligned W^{(ab)}(y,z)
&={1\over (y-z)^{\delta_{ab}}}
(X^{(ab)}(y,z)-\delta_{ab}),\quad\text{where}\\
X^{(ab)}(y,z):&=\\
\epsilon(\delta_a,\delta_b)&e^{\delta_a-\delta_b}y^{\delta_a}z^{-\delta_b}
\exp(\sum_{k=1}^{\infty}(t_k^{(a)}y^k-t_k^{(b)}z^k))
\exp(-\sum_{k=1}^{\infty}(
{\partial\over \partial t_k^{(a)}}{y^{-k}\over k}-
{\partial\over \partial t_k^{(b)}}{z^{-k}\over k})
)\endaligned \tag {3.11}$$

The fields for which $\ell=0$, give the
vertex operator realization of the homogeneous
realization of $\hat{gl}_s$, which was first found by
Frenkel and Kac [FK] and independently by Segal [Se](see also [TV] for more
details).

Using the isomorphism $\sigma$ we can reformulate the KP hierarchy
(1.9) in the bosonic picture.
We start by observing that (1.9) can be rewritten as follows:
$$\text{Res}_{z=0}\ dz ( \sum^{s}_{j=1} \psi^{+(j)}(z)\tau
\otimes \psi^{-(j)}(z)\tau ) = 0,\ \tau \in F^{(0)}.
\tag{3.12}$$
Notice that for $\tau \in F^{(0)},\ \sigma (\tau) = \sum_{\gamma \in Q}
\tau_{\gamma}(t)e^{\gamma}$.
 Here and further  we write $\tau_{\gamma}(t)e^{\gamma}$ for
$\tau_{\gamma} (t)\otimes
e^{\gamma}$.  Using  (3.8), equation (3.12) turns under $\sigma
\otimes \sigma :\ F \otimes F \overset\,\,\sim\to\longrightarrow
{\Bbb C}[t,t^{\prime }]
\otimes ({\Bbb C}[L^{}] \otimes {\Bbb C}[L^{ \prime}])$ into the
following set of equations;
 for all $\alpha ,\beta \in L$ such that $(\alpha
|\delta ) = -(\beta |\delta ) = 1$ we have:
$$\aligned
&\text{Res}_{z=0} ( dz
 \sum^{s}_{j=1} \varepsilon (\delta_{j}, \alpha-\beta)
z^{(\delta_{j}|\alpha - \beta - 2\delta_{j})}  \\
 &\times \exp
(\sum^{\infty}_{k=1} (t^{(j)}_{k} - t^{(j)^{
\prime}}_{k})z^{k})
\exp (-\sum^{\infty}_{k=1} (\frac{\partial}{\partial t^{(j)}_{k}}
 - \frac {\partial}{\partial t^{(j)^{
\prime}}_{k}})\frac{z^{-k}}{k}) \\
& \tau_{\alpha -
\delta_{j}}(t)(e^{\alpha})\tau_{\beta + \delta_{j}}(t^{
\prime})(e^{\beta})^{\prime})
= 0  . \endaligned \tag{3.13}$$
Notice that if $s=2$, the set of equations (3.13) are equivalent
(for more general $\tau$) to the Toda lattice hierarchy of Ueno and Takasaki
[UT].
For this reason,
{\it we assume from now on that $\tau=\sum \tau_\alpha e^\alpha$ is any
solution
of} (3.13). Hence the results of this paper also hold for an
(alternative definition) of the Toda lattice hierarchy.

In order to define the equations (3.13) in terms of formal
pseudo--differential operators it will be convenient to replace
$t_k^{(j)}$ by $t_k^{(j)}+\delta_{k,1}x$ and to introduce the notations
$$\aligned &\xi^{(j)}(t,z)=\sum_{i=1}^{\infty}t^{(j)}_iz^i,\quad
 \xi^{(j)}(x,t,z)=zx+\xi^{(j)}(t,z)\\ &\text{ and}\qquad \eta^{(j)}(t,z)=
\sum_{i=1}^{\infty}{\partial\over\partial t^{(j)}_i}{z^{-i}\over
i.}\endaligned$$
Next we replace $\alpha$ resp. $\beta$ by $\alpha+\delta_i$ and
$\beta-\delta_k$
then for all $\alpha,\beta\in Q$ and $1\le i,k\le s$ (3.13) turns into
$$\aligned
&\text{Res}_{z=0}  dz
 (\sum^{s}_{j=1} \varepsilon (\delta_{j}, \alpha+\delta_i-\beta+\delta_k)
z^{(\delta_{j}|\alpha - \beta +\delta_i+\delta_k-
2\delta_{j})}e^{\xi^{(j)}(x,t,z)
-\xi^{(j)}(x',t',z)}\\
&\times e^{-\eta^{(j)}(t,z)+\eta^{(j)}(t',z)}\tau_{\alpha +\delta_i-
\delta_{j}}(x,t)
(e^{\alpha+\delta_j})\tau_{\beta + \delta_{j}-\delta_k}
(x',t^\prime )(e^{\beta-\delta_k})^\prime)
= 0  . \endaligned \tag{3.14}$$

\vskip 10pt
\subheading{\S 4. The algebra of formal pseudo-differential operators and the
$s$-component KP hierarchy as a dynamical system }

\vskip 10pt
\noindent We proceed now to rewrite the formulation (3.14) of the
$s$-component KP hierarchy in terms of formal pseudo-differential
operators, generalizing the results of [DJKM1-3]. For more details
see [KV].
For each $\alpha \in \ \text{supp}\ \tau :=\{\alpha\in Q| \tau=\sum_{\alpha\in
Q}\tau_{\alpha}e^{\alpha}, \tau_{\alpha}\ne 0\}$ we define the (matrix
valued) functions
$$\Psi^\pm(\alpha,z)\equiv \Psi^{\pm} (\tau_\alpha ,x,t,z)
= (\Psi^{\pm}_{ij}(\tau_\alpha ,x,t,z))^{s}_{i,j=1}
\tag{4.1}$$
as follows:
$$\aligned
\Psi^{\pm}_{ij}(\tau_\alpha ,x,t,z) \overset{\text{def}}\to{=}&
\varepsilon (\delta_{j} , \alpha + \delta_{i})
 z^{(\delta_{j}|\pm \alpha + \delta_{i}-\delta_{j})}e^{\pm \xi^{(j)}(x,t,z)}
e^{\mp \eta^{(j)}(t,z)} \tau_{\alpha  \pm
(\delta_{i}-\delta_{j})}  (x,t)/\tau_{\alpha}(x,t)\\
=&\varepsilon (\delta_{j} , \alpha + \delta_{i})
 z^{(\delta_{j}|\pm \alpha + \delta_{i}-\delta_{j})}
\tau_{\alpha  \pm
(\delta_{i}-\delta_{j})}  (x,t^{(k)}-\delta_{jk}[z^{-1}])/\tau_{\alpha}(x,t)
e^{\pm \xi^{(j)}(x,t,z)},
\endaligned\tag{4.2}
$$
where $[w]=(w,{w^2\over 2},{w^3\over 3},\ldots)$
It is easy to see that equation (3.14) is equivalent to the
following bilinear identity:
$$Res_{z=0}\Psi^{+}(\tau_\alpha ,x,t,z)\ ^{t}\Psi^{-}(\tau_\beta
,x^{\prime},t',z)dz = 0\
\text{for all}\ \alpha ,\beta \in Q. \tag{4.3}$$
Define $s \times s$ matrices $S^{\pm (m)} (\tau_\alpha ,x,t)$ by the
following generating series (cf. (4.2)):
$$\aligned
\sum^{\infty}_{m=0}
S^{\pm (m)}_{ij} (\tau_\alpha ,x,t)(\pm z)^{-m}
=& \varepsilon_{ji}z^{\delta_{ij}-1} e^{\mp\eta(t,z)}
\tau_{\alpha \pm
(\delta_i-\delta_j)} (x,t))/\tau_{\alpha} (x,t)\\
=&\varepsilon_{ji}z^{\delta_{ij}-1}
\tau_{\alpha \pm
(\delta_i-\delta_j)} (x,t^{(k)}-\delta_{jk}[z^{-1}]))/\tau_{\alpha}
(x,t).
\endaligned
\tag{4.4}
$$
We see from (4.2) that $\Psi^{\pm}(\tau_\alpha ,x,t,z)$ can be written in the
following form:
$$\Psi^{\pm}(\tau_\alpha ,x,t,z) = (\sum^{\infty}_{m=0}
S^{\pm (m)}(\tau_\alpha ,x,t)R^{\pm}(\alpha ,\pm z)(\pm
z)^{-m})e^{\pm \xi (x,t,z)}, \tag{4.5}$$
where
$$R^{\pm}(\alpha ,z) = \sum^{s}_{i=1}
\varepsilon (\delta_{i}, \alpha ) e_{ii} (\pm z)^{\pm
(\delta_{i}|\alpha )}. \tag{4.6}$$
as before  $e_{ij}$ stands for the $s \times s$ matrix whose
$(i,j)$ entry is $1$ and all other entries are zero. Let
$$\partial = \frac{\partial}{\partial x},$$ we can now rewrite
$\Psi^{\pm}(\tau_\alpha ,x,t,z)$ in
terms of formal pseudo-differential operators,
define
$$\aligned &e^{\pm \xi(t,\pm\partial)}=\sum_{j=1}^s e^{\pm
\xi^{(j)}(t,\pm\partial)}
e_{jj}\\
&P^{\pm}(\alpha ) \equiv P^{\pm} (\tau_\alpha ,x,t,\partial ) =
I_{s} + \sum^{\infty}_{m=1} S^{\pm (m)} (\tau_\alpha ,x,t)\partial^{-m},\\
&R^{\pm}(\alpha ) \equiv R^{\pm}(\alpha ,\partial)\ \text{and}\
W^{\pm}(\alpha)\equiv
W^{\pm}(\tau_\alpha,x,t,z)=P^{\pm}(\alpha)R^{\pm}(\alpha)e^{\pm
\xi(t,\pm\partial)}
\endaligned \tag{4.7}$$
then:
$$\Psi^{\pm}(\tau_\alpha ,x,t,z) =W^{\pm}(\alpha)e^{\pm zx}= P^{\pm } (\alpha )
R^{\pm}(\alpha)e^{\pm\xi(t,\pm\partial)}e^{\pm zx}. \tag{4.8}$$

As usual one denotes the differential part of $P(x,t,\partial )$ by
$P_{+}(x,t,\partial ) = \sum_{j\ge 0} P_{j}(x,t) \partial^{j},$
and writes $P_{-} = P-P_{+}$. The  linear  anti-involution   $*$ is defined by
the
following formula:
$$(\sum_{j} P_{j}\partial^{j})^{*} = \sum_{j} (-\partial )^{j}
\circ ^{t}\! P_{j}. \tag{4.9}$$
Here and further $^{t}P$ stands for the transpose of the matrix $P$.
Then one has the following fundamental lemma:

\proclaim{Lemma 4.1} Let $P,Q$ be two formal pseudo--differential
operators, then $$(PQ^{*})_-=\sum_{i<0}R_i\partial^i$$
if and only if
$$\text{Res}_{z=0} dz (P(x,\partial) e^{z x})  \ ^{t}
(Q(x^{\prime},\partial^{\prime}) e^{-z x^{\prime}})  = \sum_{i<0}
R_i(x){(x-x')^{-i-1}\over (-i-1)!}.
$$
\endproclaim
\demo {Proof}
Let $y=x-x'$, $P(x,z)=\sum P_i(x)z^i$ and $Q(x,z)=\sum Q_i(x)(-z)^i$,
then
$$\aligned &\text{Res}_{z=0} dz(P(x,\partial) e^{z x})  \ ^{t}
(Q(x^{\prime},\partial^{\prime}) e^{-z x^{\prime}})
 =\\
&\text{Res}_{z=0} dz P(x,z)\sum_{k\ge 0}{(-1)^k\over k!}
{\partial^{k\ t}Q(x,-z)\over \partial x^k}y^ke^{zy}=\\
&\text{Res}_{z=0} dz \sum_{k,\ell\ge 0,i,j}{(-1)^k\over k!}
P_i(x){\partial^{k\ t}Q_j(x)\over \partial x^k}{y^{k+\ell}\over\ell !}
z^{i+j+\ell}=\\
&\sum_{k\ge 0,i+j\le -1}{(-1)^k\over k!(-i-j-1)!}
P_i(x){\partial^{k\ t}Q_j(x)\over \partial x^k}y^{k-i-j-1}=\\
&\sum_{k\ge 0,i+j\le -1}{i+j\choose k}
P_i(x){\partial^{k\ t}Q_j(x)\over \partial x^k}{y^{k-i-j-1}\over
(k-i-j-1)!}.
\endaligned\tag{4.10}
$$
Next we calculate
$$\aligned
(P(x,\partial)
Q^{*}(x,\partial))_-&=(\sum_{i,j}P_i(x)\partial^{i+j\ t}Q_j(x))_-\\
&=(\sum_{k\ge 0,i,j}{i+j\choose k}P_i(x){\partial^{k\ t}Q_j(x)
\over\partial x^k}\partial^{i+j-k})_-\\
&=(\sum_{k\ge 0,i+j-k<0}{i+j\choose k}P_i(x){\partial^{k\ t}Q_j(x)
\over\partial x^k}\partial^{i+j-k})_-\\
&=\sum_{k\ge 0,i+j\le -1}{i+j\choose k}
P_i(x){\partial^{k\ t}Q_j(x)\over \partial x^k}\partial^{i+j-k},
\endaligned \tag{4.11}$$
here we have used the fact that ${i+j\choose k}=0$ if $i+j>0$ and $i+j-k<0$.
Now comparing (4.10) and (4.11) gives the desired result.
$\qquad\square$

\enddemo

Using this Lemma one deduces the following
$$(W^+(\tau_\alpha,x,t,\partial)W^-(\tau_\beta,x,t',\partial))_-=0.\tag{4.12}$$
By putting $t=t'$, one proves in a similar way as in [KV] that
given $\beta \in \text{supp}\ \tau$, all the
pseudo-differential operators $P^{\pm} (\alpha )$, $\alpha \in \text{supp}\
\tau$, are
completely determined by $P^+(\beta )$ from the following equations
$$\align
&R^{-}(\alpha )^{-1} = R^{+}(\alpha
)^{*} ,
\tag{4.13}\\
&P^{-}(\alpha  ) = (P^{+}(\alpha  )^{*})^{-1} ,
\tag{4.14}\\
&(P^{+}(\alpha)R^{+}(\alpha - \beta)P^{+}(\beta)^{-1})_{-} = 0\
\text{for all}\ \alpha ,\beta \in \text{supp}\ \tau . \tag{4.15}
\endalign$$
This and the above lemma can be used to prove the following
proposition which will be crucial later on.
Adler, Shiota and van Moerbeke stated this proposition in
the 1--component case
[ASV2].
\proclaim{Proposition 4.2} Let $\Psi^\pm(\alpha,x,t,z)$ satisfy the
bilinear
identity (4.3) for $\beta=\alpha$ and let $Q(x,t,\partial)$ be an
arbitrary
pseudo--differential operator. Then $Q$ is a differential operator
if and only if
$$\text{Res}_{z=0}dz
Q(x,t,\partial)\Psi^+(\tau_\alpha,x,t,z)\Psi^-(\tau_\alpha,x',t',z)=0\tag{4.16}
$$
\endproclaim
\demo{Proof}
Suppose that $Q$ is a differential operator, then since by lemma 4.1
\break
 $W^+(\tau_\alpha,x,t,\partial)W^-(\tau_\alpha,x,t',\partial)$
is a differential operator
$$(Q(x,t,\partial)W^+(\tau_\alpha,x,t,\partial)W^-(\tau_\alpha,x,t',\partial))_-=0.\tag{4.17}$$
Conversely, suppose (4.16) holds, then again by Lemma , (4.17) holds. Now put
$t=t'$, and use (4.13-14), then one deduces that $Q(x,t,\partial)_-=0$.$
\qquad\square$
\enddemo
In [KV] Victor Kac and the author also showed
 the following
\proclaim{Proposition 4.3}  Consider $\Psi^{+}(\tau_\alpha ,x,t,z)$ and
 $\Psi^{-}(\tau_\alpha ,x,t,z),$ $\alpha \in Q$, of the
form (4.8), then the
bilinear identity (4.3) for all $\alpha ,\beta \in \ \text{supp}\
\tau$ is equivalent to the Sato equation:
$$
\frac{\partial P^+(\alpha)}{\partial t^{(j)}_{k}} = -(P^+(\alpha)e_{jj}
\partial^{k}  P^+(\alpha)^{-1})_{-}  P^+(\alpha), \tag{4.18}$$
for each $\alpha \in\text{supp}\ \tau$ and the matching conditions (4.13-15)
for
all $\alpha ,\beta \in \ \text{supp} \ \tau$.
\endproclaim
\noindent As a consequence of (4.18) one obtains for each
$W^+(\alpha)$:
$$\aligned \frac{\partial W^+(\alpha) }{\partial t^{(j)}_{k}} &=
(P^+(\alpha)e_{jj}
\partial^{k}  P^+(\alpha)^{-1})_{+}W^+(\alpha)
\\
&=(W^+(\alpha)e_{jj}\partial^kW^+(\alpha)^{-1})_+W^+(\alpha)\endaligned$$
  Fix $\tau$ and $\alpha ,\beta\in Q$, introduce the following formal
pseudo-differential
operators $L(\alpha)$, $ \Gamma$, $M(\alpha)$, $N(\alpha,\beta)$,
$\Delta^{(ij)}$, $ C^{(ij)}(\alpha )$  and
differential operators  $B^{(ij)}_{m}(\alpha)$:
$$\aligned
L(\alpha )
  & = W^+(\alpha)\partial W^+(\alpha)^{-1}=
P^{+}(\alpha)  \partial  P^{+}(\alpha)^{-1}, \\
\Gamma&=
x+\sum_{a=0}^s \sum_{k=0}^\infty kt^{(a)}_k\partial^{k-1}e_{aa}\\
M(\alpha)
&=W^+(\alpha)xW^+(\alpha)^{-1}=
P^{+}(\alpha)\Gamma
   P^{+}(\alpha)^{-1},\\
N(\alpha,\beta)&=W^+(\alpha)W^+(\beta)^{-1}
=P^+(\alpha)R^+(\alpha-\beta)P^+(\beta)^{-1}\\
\Delta^{(ij)}&=e^{\xi(t,\partial)}e_{ij}e^{-\xi(t,\partial)}\\
C^{(ij)}(\alpha ) &=
 W^+(\alpha)e_{ij} W^+(\alpha)^{-1}=
P^{+}(\alpha)\Delta^{(ij)}P^{+}(\alpha)^{-1}, \\
B^{(ij)}_{m}(\alpha) &=
( W^+(\alpha)e_{ij}\partial^mW^+(\alpha)^{-1})_+=
(P^{+}(\alpha)e_{ij} \partial^{m}
P^{+}(\alpha)^{-1})_{+}.
\endaligned \tag{4.20}
$$
Here we write $x$ for $xI_s$. Denote by
$C^{(i)}(\alpha)=C^{(ii)}(\alpha)$
and $B^{(i)}_m(\alpha)=B^{(ii)}_m(\alpha)$. Then
$$\aligned
&[L(\alpha),M(\alpha)]=I_s,\
\sum^{s}_{i=1} C^{(i)}(\alpha) = I_{s}, \
C^{(ij)}(\alpha)L(\alpha) = L(\alpha)C^{(ij)}(\alpha),\\
&C^{(ij)}(\alpha)C^{(k\ell)}(\alpha) = \delta_{jk}
C^{(i\ell)}(\alpha),\
L(\alpha)N(\alpha,\beta)=N(\alpha,\beta)L(\beta),\\
&M(\alpha)N(\alpha,\beta)=N(\alpha,\beta)M(\beta),\
C^{(ij)}(\alpha)N(\alpha,\beta)=N(\alpha,\beta)C^{(ij)}(\beta),\\
&N(\alpha,\beta)N(\beta,\gamma)=N(\alpha,\gamma).
\endaligned
 \tag{4.21}$$
\proclaim{Remark 4.4}(i) It is our purpose to describe the general
operators
$$Y^{(ab,\ell+1)}_k(\alpha,\beta)=W^+(\alpha)x^\ell\partial^{k+\ell}e_{ab}
W^+(\beta)^{-1}.$$
One can express them in the operators defined in (4.20), viz.,
$$Y^{(ab,\ell+1)}_k(\alpha,\beta)=M(\alpha)^\ell L(\alpha)^{k+\ell}
C^{(ab)}(\alpha)N(\alpha,\beta).$$
(ii) Notice that (4.14) is equivalent to $N(\alpha,\beta)_-=0$,
so from now on we will assume that $N(\alpha,\beta)$ is a differential
operator.
\endproclaim
\proclaim{Proposition 4.5} If for every $\alpha, \beta \in Q$
the
formal pseudo-differential operators $L(\alpha)$,
 $M(\alpha)$, $
C^{(ij)}(\alpha)$  and the differential operators $N(\alpha,\beta)$ satisfy
conditions (4.21) and if the
equations
$$
\cases
L(\alpha)P^+(\alpha)R^+(\alpha) = P^+(\alpha)R^+(\alpha)\partial \quad(\text{or
equivalently}\ L(\alpha)P^+(\alpha)=P^+(\alpha)\partial)&\ \\
M(\alpha)P^+(\alpha)R^+(\alpha)=P^+(\alpha)R^+(\alpha)\Gamma
&\ \\
N(\alpha,\beta)P^+(\beta)R^+(\beta)=P^+(\alpha)R^+(\alpha)&\ \\
C^{(ij)}(\alpha)P^+(\alpha)R^+(\alpha) = P^+(\alpha)R^+(\alpha)\Delta^{(ij)} &\
\\
\displaystyle{\frac{\partial P^+(\alpha)}{\partial t^{(j)}_{k}} =
-(C^{(j)}(\alpha)L(\alpha)^{k})_{-} P^+(\alpha)}&\
\endcases
\tag{4.22}$$
 have a solution $P^+(\alpha)$ of the form (4.7), then the
differential operators $B^{(j)}_{k}(\alpha)$
satisfies the following
conditions:
$$\cases \displaystyle{\frac{\partial L(\alpha)}{\partial t^{(j)}_{k}} =
[B^{(j)}_{k}(\alpha),L(\alpha)],}  & \   \\
\displaystyle{\frac{\partial M(\alpha)}{\partial t^{(j)}_{k}} =
[B^{(j)}_{k}(\alpha),M(\alpha)],}  & \   \\
\displaystyle{\frac{\partial N(\alpha,\beta)}{\partial t^{(j)}_{k}} =
B^{(j)}_{k}(\alpha)N(\alpha,\beta)-N(\alpha,\beta)B^{(j)}_{k}(\beta),}&\
\\
\displaystyle{\frac{\partial C^{(i\ell)}(\alpha)}{\partial t^{(j)}_{k}} =
[B^{(j)}_{k}(\alpha),C^{(i\ell)}(\alpha)],} & \
\  \endcases. \tag{4.23}$$
\endproclaim
\noindent Now (4.23) implies that
$$\frac{\partial Y^{(ab,\ell+1)}_m(\alpha, \beta)}{\partial t^{(j)}_{k}} =
B^{(j)}_{k}(\alpha)Y^{(ab,\ell+1)}_m(\alpha, \beta)
-Y^{(ab,\ell+1)}_m(\alpha, \beta)B^{(j)}_{k}(\beta)$$
Notice that the conditions (4.22) are equivalent to one of the following
conditions:
$$
\cases L(\alpha)W^+(\alpha)= W^+(\alpha)\partial, &\ \\
M(\alpha)W^+(\alpha)=W^+(\alpha)x, &\ \\
N(\alpha,\beta)W^+(\beta)=W^+(\alpha),&\ \\
C^{(ij)}(\alpha)W^+(\alpha) = W^+(\alpha)e_{ij}, &\ \\
\displaystyle{\frac{\partial W^+(\alpha)}{\partial t^{(j)}_{k}}
= B^{(j)}_k(\alpha) W^+(\alpha),}
\endcases
\qquad
\cases L(\alpha)\Psi^+(\alpha,z)= z\Psi^+(\alpha,z),&\ \\
M(\alpha)\Psi^+(\alpha,z)={\partial\Psi^+(\alpha,z)\over\partial z},
&\ \\
N(\alpha,\beta)\Psi^+(\beta,z)=\Psi^+(\alpha,z),&\ \\
C^{(ij)}(\alpha)\Psi^+(\alpha,z) = \Psi^+(\alpha,z) e_{ij}, &\ \\
\displaystyle{\frac{\partial \Psi^+(\alpha,z)}{\partial t^{(j)}_{k}}
= B^{(j)}_k(\alpha) \Psi^+(\alpha,z).}
\endcases
\tag{4.24}$$

\subheading{\S5 The Adler--Shiota--van Moerbeke formula}
\vskip 10pt
\noindent
Fix $\alpha, \beta\in Q$ and recall
$$Y^{(ab,\ell+1)}_k(\alpha,\beta)\equiv
Y^{(ab,\ell+1)}_k(\tau_\alpha,\tau_\beta)
=W^+(\alpha)x^\ell\partial^{k+\ell}e_{ab}
W^+(\beta)^{-1},\tag{5.1}$$
then define
$$\aligned
Y^{(ab)}(\alpha,\beta,y,w)&\equiv Y^{(ab)}(\tau_\alpha,\tau_\beta,y,w)=
\sum_{\ell=0}^\infty {(y-w)^\ell\over\ell !}\sum_{k\in \Bbb Z}
w^{-k-\ell-1}Y^{(ab,\ell+1)}_k(\alpha,\beta)\\
&=\sum_{\ell=0}^\infty {(y-w)^\ell\over\ell !}\sum_{k\in \Bbb Z}
w^{-k-\ell-1}M(\alpha)^\ell L(\alpha)^{k+\ell}C^{(ab)}(\alpha)N(\alpha,\beta).
\endaligned\tag{5.2}$$
We write $Y^{(ab,\ell+1)}_k(\alpha)$ and
$Y^{(ab)}(\alpha,y,w)\equiv Y^{(ab)}(\tau_\alpha,y,w)$ for respectively
$Y^{(ab,\ell+1)}_k(\alpha,\alpha)$,
$Y^{(ab)}(\alpha,\alpha,y,w)$,
then
$$Y^{(ab)}(\alpha,y,w)=\sum_{\ell=0}^\infty {(y-w)^\ell\over\ell !}\sum_{k\in
\Bbb Z}
w^{-k-\ell-1}M(\alpha)^\ell L(\alpha)^{k+\ell}C^{(ab)}(\alpha).$$
One deduces  the following
\proclaim{Proposition 5.1}
$$Y^{(ab)}(\alpha,\beta,y,w)_-=\Psi^+(\alpha,y)e_{ab}\partial^{-1}\,^t\Psi^-(\beta,
w)\tag{5.3}$$
\endproclaim
\demo{Proof}
First, notice that
$$\aligned
(W^+(\alpha)x^\ell\partial^{k+\ell}e_{ab}W^{+}(\beta)^{-1})_-
&=\sum_{j=1}^\infty\partial^{-j}\text{Res}_\partial\partial^{j-1}
W^+(\alpha)x^\ell\partial^{k+\ell}e_{ab}W^{+}(\beta)^{-1}\\
&=\sum_{j=1}^\infty\partial^{-j}\text{Res}_{z=0}dz (\partial^{j-1}
W^+(\alpha)x^\ell\partial^{k+\ell}e^{zx})e_{ab}\,^t(W^-(\beta)e^{-zx})\\
&=\text{Res}_{z=0}dz\sum_{j=1}^\infty\partial^{-j}[({\partial\over\partial
x})^{j-1}(z^{k+\ell}{\partial^\ell \Psi^+(\alpha,z)\over\partial
z^\ell})]e_{ab}\,^t\Psi^-(\beta,z)\\
&=\text{Res}_{z=0}dz z^{k+\ell}{\partial^\ell \Psi^+(\alpha,z)\over\partial
z^\ell} e_{ab}\partial^{-1}\,^t\Psi^-(\beta,z)
\endaligned
$$
Hence
$$\aligned
Y^{(ab)}(\alpha,\beta,y,w)_-&=\sum_{\ell=0}^\infty {(y-w)^\ell\over\ell
!}\sum_{k\in \Bbb Z}
w^{-k-\ell-1}\text{Res}_{z=0}dz z^{k+\ell}{\partial^\ell
\Psi^+(\alpha,z)\over\partial
z^\ell} e_{ab}\partial^{-1}\,^t\Psi^-(\beta,z)\\
&=\sum_{\ell=0}^\infty {(y-w)^\ell\over\ell !}
{\partial^\ell \Psi^+(\alpha,w)\over\partial
w^\ell} e_{ab}\partial^{-1}\,^t\Psi^-(\beta,w)\\
&=\Psi^+(\alpha,y) e_{ab}\partial^{-1}\,^t\Psi^-(\beta,w)\qquad\square
\endaligned
$$
\enddemo
\noindent Proposition 5.1 was obtained in the 1--component case by
Dickey [D].

Next we calculate
$$\aligned
Y^{(ab)}(\alpha,\beta,y,w)\Psi^+(\beta,z)
&=\sum_{\ell=0}^\infty {(y-w)^\ell\over\ell !}\sum_{k\in \Bbb Z}
w^{-k-\ell-1}W^+(\alpha)x^\ell\partial^{k+\ell}e_{ab}e^{zx}\\
&=W^+(\alpha)\delta(w,z)e^{(y-w){\partial\over\partial
z}}e^{zx}e_{ab}\\
&=W^+(\alpha)\delta(w,z)e^{(z+y-w)x}e_{ab}\\
&=W^+(\alpha)\delta(w,z)e^{yx}e_{ab}\\
&=\delta(w,z)\Psi^+(\alpha,y)e_{ab},
\endaligned\tag{5.4}
$$
where $\delta(w,z)=\sum_{n\in\Bbb Z}w^{-n}z^{n-1}$.

Define $$\aligned
\Bbb X^{(ab)}(y,w)&=X^{(ab)}(y,w)e^{(y-w)x},\\
\Bbb W^{(ab)}(y,w)&=\sum_{\ell=0}^\infty {(y-w)^\ell\over \ell !}
\sum_{k\in \Bbb Z}\Bbb W_k^{(ab,\ell+1)} w^{-k-\ell-1}\\
&=(y-w)^{-\delta_{ab}}(\Bbb X^{(ab)}(y,w)-\delta_{ab}),\endaligned
\tag{5.5}$$
then
$$\Bbb W^{(ab)}(y,w)=:\psi^{+(a)}(y)\psi^{-(b)}(w):e^{(y-w)x}.$$
It is straightforward that the ${\Bbb W}_k^{(ab,\ell+1)}$ have the
same commutation relations as the $ W_k^{(ab,\ell+1)}$, since we have
only replaced  all $t^{(j)}_1$ by $t^{(j)}_1+x$ and kept
$\partial\over\partial t^{(j)}_1$ unchanged in the vertex operator
(3.11) of $W^{(ab)}(y,w)$ ($\partial\over\partial x$ does not appear
in this expression).

One has the following
\proclaim {Lemma 5.2}
$$\Bbb
X^{(ab)}(y,w)\psi^{+(j)}(z)e^{zx}=\delta_{bj}(y-w)^{\delta_{ab}}
\delta(w,z)\psi^{+(a)}(y)e^{yx}+\psi^{+(j)}(z)e^{zx}\Bbb
X^{(ab)}(y,w).$$
\endproclaim
\demo{Proof}
Let $\gamma\in Q$, we calculate
$$\aligned \Bbb
X^{(ab)}&(y,w)\psi^{+(j)}(z)e^{zx}\tau_\gamma(x,t)e^\gamma=\\
&\epsilon(\delta_a,\delta_b)\epsilon(\delta_a-\delta_b,\gamma+\delta_j)
\epsilon(\delta_j,\gamma)y^{(\delta_a|\gamma+\delta_j)}
w^{-(\delta_b|\gamma+\delta_j)}z^{(\delta_j|\gamma)}
(1-{z\over y})^{\delta_{aj}}(1-{z\over w})^{-\delta_{bj}}\\
&\qquad\times e^{\xi^{(a)}(x,t,y)-\xi^{(b)}(x,t,w)+\xi^{(j)}(x,t,z)}
e^{-\eta^{(a)}(t,y)+\eta^{(b)}(t,w)-\eta^{(j)}(t,z)}\tau_\gamma
e^{\gamma+\delta_j+\delta_a-\delta_b}
\endaligned$$
Now use the fact that
$(1-{z\over w})^{-1}=w\delta(w,z)-{w\over z}(1-{w\over z})^{-1}$,
$1-{z\over y}=-{z\over y}(1-{y\over z})$ and that
$\epsilon(\delta_m,\delta_n)\epsilon(\delta_n,\delta_m)=-(-1)^{\delta_{mn}}$,
then
$$\aligned \Bbb
X^{(ab)}&(y,w)\psi^{+(j)}(z)e^{zx}\tau_\gamma(x,t)e^\gamma\\
&=\delta_{bj}
\epsilon(\delta_a,\delta_b)\epsilon(\delta_a-\delta_b,\gamma+\delta_b)
\epsilon(\delta_b,\gamma)(y-z)^{\delta_{ab}}\delta(w,z)
y^{(\delta_a|\gamma)}
w^{-(\delta_b|\gamma)}z^{(\delta_b|\gamma)}\\
&\qquad\times e^{\xi^{(a)}(x,t,y)-\xi^{(b)}(x,t,w)+\xi^{(b)}(x,t,z)}
e^{-\eta^{(a)}(t,y)+\eta^{(b)}(t,w)-\eta^{(b)}(t,z)}\tau_\gamma
e^{\gamma+\delta_a}\\
&\quad+\epsilon(\delta_a,\delta_b)\epsilon(\delta_a-\delta_b,\gamma+\delta_j)
\epsilon(\delta_j,\gamma)(-)^{\delta_{aj}}(-)^{\delta_{bj}}
y^{(\delta_a|\gamma)}
w^{-(\delta_b|\gamma)}z^{(\delta_j|\gamma+\delta_a-\delta_b)}\\
&\qquad\times e^{\xi^{(j)}(x,t,z)}e^{\eta^{(j)}(t,z)}
e^{\xi^{(a)}(x,t,y)-\xi^{(b)}(x,t,w)}
e^{-\eta^{(a)}(t,y)+\eta^{(b)}(t,w)}\tau_\gamma
e^{\gamma+\delta_j+\delta_a-\delta_b}\\
&=\{ \delta_{bj}(y-w)^{\delta_{ab}}
\delta(w,z)\psi^{+(a)}(y)e^{yx}+
+\psi^{+(j)}(z)e^{zx}\Bbb
X^{(ab)}(y,w)\}\tau_\gamma e^\gamma.\qquad\square
\endaligned
$$\enddemo

Recall the bilinear identity (3.14) with $\alpha$ replaced by
$\alpha+\delta_b-\delta_a$ in a slightly different version, viz.,
$$\aligned \text{Res}_{z=0}dz&\sum_{j=1}^s\psi^{+(j)}(z)e^{zx}
\tau_{\alpha+\delta_i+\delta_b-\delta_j-\delta_a}(x,t)
e^{\alpha+\delta_i+\delta_b-\delta_j-\delta_a}\\
&\psi^{-(j)'}(z)e^{-zx'}
\tau_{\beta+\delta_j-\delta_k}(x',t')(e^{\beta+\delta_j-\delta_k})' =0.
\endaligned
$$
Let $\Bbb X^{(ab)}(y,w)$ act on this identity, then using Lemma 5.2
one obtains:
$$\aligned
&\text{Res}_{z=0}dz\{ (y-w)^{\delta_{ab}}\delta(w,z)
\psi^{+(a)}(y)e^{zx}
\tau_{\alpha+\delta_i-\delta_a}(x,t)
e^{\alpha+\delta_i-\delta_a}\psi^{-(a)'}(z)\\
&+\sum_{j=1}^s\psi^{+(j)}(z)e^{zx}\Bbb X^{(ab)}(y,w)
\tau_{\alpha+\delta_i+\delta_b-\delta_j-\delta_a}(x,t)
e^{\alpha+\delta_i+\delta_b-\delta_j-\delta_a}\psi^{-(j)'}(z)\} \\
&\times e^{-zx'}
\tau_{\beta+\delta_j-\delta_k}(x',t')(e^{\beta+\delta_j-\delta_k})'=0.\endaligned
\tag{5.6}$$
Now divide by $\tau_\alpha(x,t)\tau_\beta(x',t')$ and remove the factors
$e^{\alpha+\delta_i}$ and $(e^{\beta-\delta_k})'$. Notice that by
doing this, the action of $\Bbb X^{(ab)}(y,w)$ is no longer
well--defined, for that reason we introduce  ${\hat X}^{(ab)}(y,w)$
by
$$\aligned
\hat X^{(ab)}&(y,w)\tau_\gamma(x,t)\\
&=\epsilon(\delta_a,\delta_b)
\epsilon(\delta_a-\delta_b,\gamma)y^{(\delta_a|\gamma)}w^{-(\delta_b|\gamma)}
e^{\xi^{(a)}(x,t,y)-\xi^{(b)}(x,t,w)}e^{-\eta^{(a)}(t,y)+\eta^{(b)}(t,w)}
\tau_\gamma(x,t).\endaligned$$ and
$$\aligned
{\hat W}^{(ab)}(y,w)&=\sum_{\ell=0}^\infty {(y-w)^\ell\over\ell
!}\sum_{k\in\Bbb Z} {\hat W}_k^{(ab,\ell+1)}w^{-k-\ell-1}\\
&=(y-w)^{-\delta_{ab}}(\hat X^{(ab)}(y,w)-\delta_{ab}).\endaligned
\tag{5.7}$$
Then (5.6) turns into
$$\aligned
\text{Res}_{z=0}dz&\{ (y-w)^{\delta_{ab}}\delta(w,z)
\Psi^{+}_{ia}(\alpha,y)\Psi^{-}_{kb}(\beta,z)'\\
&+\sum_{j=1}^s
e^{-\eta^{(j)}(t,z)}\left ({{\hat X}^{(ab)}(y,w)
\tau_{\alpha+\delta_i+\delta_b-\delta_j-\delta_a}(x,t)\over
\tau_{\alpha+\delta_i-\delta_j}(x,t)}\right )\Psi^{+}_{ij}(\alpha,z)
\Psi^{-}_{kj}(\beta,z)'\}=0.\endaligned$$
Using (5.4) one obtains
$$
\aligned
\text{Res}_{z=0}dz&\{ e_{ii}  ( (y-w)^{\delta_{ab}}Y^{(ab)}(\alpha,y,w)
\Psi^+(\alpha,z)\\
&+\sum_{j=1}^s
e^{-\eta^{(j)}(t,z)}\left ({{\hat X}^{(ab)}(y,w)
\tau_{\alpha+\delta_i+\delta_b-\delta_j-\delta_a}(x,t)\over
\tau_{\alpha+\delta_i-\delta_j}(x,t)}\right ))\Psi^{+}(\alpha,z)e_{jj}
\}^t\Psi^{-}(\beta,z)'=0.\endaligned$$
Now notice that
$$\aligned
e^{-\eta^{(j)}(t,z)}&\left ({{\hat X}^{(ab)}(y,w)
\tau_{\alpha+\delta_i+\delta_b-\delta_j-\delta_a}(x,t)\over
\tau_{\alpha+\delta_i-\delta_j}(x,t)}\right
)\Psi^{+}(\alpha,z)e_{jj}=
\sum_{k=0}^\infty
c_{jk}L(\alpha)^{-k}C^{(j)}(\alpha)\Psi^{+}(\alpha,z)\\
&=\{ (\sum_{k=1}^\infty
c_{jk}L(\alpha)^{-k}C^{(j)}(\alpha))_-
+{{\hat X}^{(ab)}(y,w)
\tau_{\alpha+\delta_i+\delta_b-\delta_j-\delta_a}(x,t)\over
\tau_{\alpha+\delta_i-\delta_j}(x,t)} e_{jj}\}\Psi^{+}(\alpha,z),
\endaligned
$$
hence
$$\text{Res}_{z=0}dz e_{ii} \left (
(y-w)^{\delta_{ab}}Y^{(ab)}(\alpha,y,w)+\sum_{j=1}^s\sum_{k=1}^\infty
c_{jk}L(\alpha)^{-k}C^{(j)}(\alpha)\right
)\Psi^+(\alpha,z)^t\Psi^{-}(\beta,z)'=0.$$
Now using Proposition 4.2 for  $\beta=\alpha$, one obtains
$$e_{ii}((y-w)^{\delta_{ab}}Y^{(ab)}(\alpha,y,w)_-+\sum_{j=1}^s(\sum_{k=0}^\infty
c_{jk}L(\alpha)^{-k}C^{(j)}(\alpha))_-)=0.$$
Hence
$$\aligned
-e_{ii}(y-w)^{\delta_{ab}}&Y^{(ab)}(\alpha,y,w)_-\Psi^+(\alpha,z)=\\
&e_{ii}\sum_{j=1}^s  e^{-\eta^{(j)}(t,z)}\left ({{\hat X}^{(ab)}(y,w)
\tau_{\alpha+\delta_i+\delta_b-\delta_j-\delta_a}(x,t)\over
\tau_{\alpha+\delta_i-\delta_j}(x,t)}\right )\Psi^+(\alpha,z)e_{jj}\\
&\qquad -
{{\hat X}^{(ab)}(y,w)
\tau_{\alpha+\delta_b-\delta_a}(x,t)\over
\tau_{\alpha}(x,t)} e_{jj}\Psi^{+}(\alpha,z).
\endaligned
$$
So we obtain the following generalization of the Adler--Shiota--van
Moerbeke formula
\proclaim{Theorem 5.3}
$$\aligned
&(y-w)^{\delta_{ab}}(-Y^{(ab)}(\alpha,y,w)_-\Psi^+(\alpha,z))_{ij}=\\
&\{ e^{-\eta^{(j)}(t,z)}\left ({{\hat X}^{(ab)}(y,w)
\tau_{\alpha+\delta_i+\delta_b-\delta_j-\delta_a}(x,t)\over
\tau_{\alpha+\delta_i-\delta_j}(x,t)}\right )-
{{\hat X}^{(ab)}(y,w)
\tau_{\alpha+\delta_b-\delta_a}(x,t)\over
\tau_{\alpha}(x,t)}\} \Psi^{+}_{ij}(\alpha,z).
\endaligned\tag {5.8}
$$
\endproclaim
In a similar way as in the introduction the operator
$e^{\lambda \Bbb X^{(ab)}(y,w)}=1+\lambda \Bbb X^{(ab)}(y,w)$ is an
auto--B\"acklund transformation of the $s$--component KP hierarchy (see
also [KV]). Now let
$$\sigma (x,t)=\sum_{\gamma\in
Q}\sigma_\gamma(x,t)e^\gamma=e^{\lambda \Bbb X^{(ab)}(y,w)}\sum_{\gamma\in
Q}\tau_\gamma(x,t)e^\gamma=
\sum_{\gamma\in Q}(\tau_\gamma(x,t)
+\lambda{\hat X}^{(ab)}(y,w)\tau_{\gamma+\delta_b-\delta_a})e^\gamma,$$
then (5.8) is equal to
$$\aligned
-\lambda(y-w)^{\delta_{ab}}(
Y^{(ab)}(\tau_\alpha,y,w)_-&\Psi^+(\tau_\alpha,x,t,z))_{ij}=\\
&\{ e^{-\eta^{(j)}(t,z)}\left ({\sigma_{\alpha+\delta_i-\delta_j}(x,t)
\over
\tau_{\alpha+\delta_i-\delta_j}(x,t)}\right )
-
{\sigma_\alpha(x,t)\over
\tau_{\alpha}(x,t)}\} \Psi^{+}_{ij}(\tau_\alpha,x,t,z).
\endaligned
$$
So
$$\aligned
\Psi_{ij}^+(\sigma_\alpha,x,t,z)&=
{e^{-\eta^{(j)}(t,z)}\sigma_{\alpha+\delta_i-\delta_j}(x,t)\over\sigma_{\alpha}(x,t)}
e^{\xi^{(j)}(x,t,z)}\\
&={\tau_\alpha(x,t)\over\sigma_{\alpha}(x,t)}e^{-\eta^{(j)}(t,z)}\left(
{\sigma_{\alpha+\delta_i-\delta_j}(x,t)\over\tau_{\alpha+\delta_i-\delta_j}(x,t)}
\right)\Psi_{ij}^+(\tau_\alpha,x,t,z)\\
&=\Psi_{ij}^+(\sigma_\alpha,x,t,z)+{\tau_\alpha(x,t)\over\sigma_{\alpha}(x,t)}
(e^{-\eta^{(j)}(t,z)}-1)\left(
{\sigma_{\alpha+\delta_i-\delta_j}(x,t)\over\tau_{\alpha+\delta_i-\delta_j}(x,t)}
\right)\Psi_{ij}^+(\tau_\alpha,x,t,z)\\
&=\Psi_{ij}^+(\sigma_\alpha,x,t,z)-\lambda{\tau_\alpha(x,t)\over\sigma_{\alpha}(x,t)}
(y-w)^{\delta_{ab}}\left
(Y^{(ab)}(\tau_\alpha,y,w)_-\Psi^+(\tau_\alpha,x,t,z)\right )_{ij}\\
&=\left (\left(1-\lambda{\tau_\alpha(x,t)\over\sigma_{\alpha}(x,t)}
(y-w)^{\delta_{ab}}Y^{(ab)}(\tau_\alpha,y,w)_-\right
)\Psi^+(\tau_\alpha,x,t,z)\right )_{ij}
\endaligned $$
and we obtain the following consequence of Theorem 5.3:
\proclaim{Corollary 5.4}
Let $\tau(x,t)$ be a solution of the $s$--component KP hierarchy, then
$\sigma(x,t)=e^{\lambda \Bbb X^{(ab)}(y,w)}\tau(x,t)$ is a new
solution of this hierarchy and the wave functions are related by
$$\Psi_{ij}^+(\sigma_\alpha,x,t,z)=
\left (\left(1-\lambda{\tau_\alpha(x,t)\over\sigma_{\alpha}(x,t)}
(y-w)^{\delta_{ab}}Y^{(ab)}(\tau_\alpha,y,w)_-\right
)\Psi^+(\tau_\alpha,x,t,z)\right )_{ij}$$
\endproclaim
\noindent Hence (5.8) relates this B\"acklund transformation of the
$s$--component KP hierarchy acting on the $\tau$--function to a
``B\"acklund transformation''
on the wave function.

Since the left--hand--side of (5.8) is also equal to
$$\aligned
 \{ e^{-\eta^{(j)}(t,z)}\left ({({\hat X}^{(ab)}(y,w)-\delta_{ab})
\tau_{\alpha+\delta_i+\delta_b-\delta_j-\delta_a}(x,t)\over
\tau_{\alpha+\delta_i-\delta_j}(x,t)}\right )&\ \\
-
{({\hat X}^{(ab)}(y,w)-\delta_{ab})
\tau_{\alpha+\delta_b-\delta_a}(x,t)\over
\tau_{\alpha}(x,t)}& \} \Psi^{+}_{ij}(\alpha,z),\endaligned$$
we have the following
\proclaim{Corollary 5.5}
$$\aligned (-(M&(\alpha)^\ell
L(\alpha)^{k+\ell}C^{(ab)}(\alpha))_-\Psi^+(\alpha,z))_{ij}=\\
&\{ e^{-\eta^{(j)}(t,z)}\left ({{\hat W}^{(ab,\ell+1)}_k
\tau_{\alpha+\delta_i+\delta_b-\delta_j-\delta_a}(x,t)\over
\tau_{\alpha+\delta_i-\delta_j}(x,t)}\right )-
{{\hat W}^{(ab,\ell+1)}_k
\tau_{\alpha+\delta_b-\delta_a}(x,t)\over
\tau_{\alpha}(x,t)}\} \Psi^{+}_{ij}(\alpha,z).\endaligned$$
\endproclaim
\demo{Proof} Compare in (5.8) the expansions for the vertex operators
$Y^{(ab)}(\alpha,y,w)$
as in (5.2) and for ${\hat W}^{(ab)}(y,w)$ as in (5.7).\qquad $\square$
\enddemo

As an application of Corollary 5.5 we see that if
$$\sum_{a=1}^s\sum_{\ell=0}^\infty\sum_{k\in\Bbb Z} c_{a\ell k}(M(\alpha)^\ell
L(\alpha)^{k+\ell}C^{(a)}(\alpha))_-=0,$$   one
finds that  that for any $1\le j\le s$
$$(e^{-\eta^{(j)}(t,z)}-1)\left
({\sum_{a=1}^s\sum_{\ell=0}^\infty\sum_{k\in\Bbb
Z} c_{a\ell k}
{\hat W}^{(a,\ell+1)}_k
\tau_{\alpha}(x,t)\over
\tau_{\alpha}(x,t)}\right )=0,$$
hence
$$\sum_{a=1}^s\sum_{\ell=0}^\infty\sum_{k\in\Bbb
Z} c_{a\ell k}
\Bbb W^{(a,\ell+1)}_k
\tau_{\alpha}(x,t)e^{\alpha}
=\text{constant}\,\tau_\alpha(x,t)e^\alpha.$$
Thus Corollary 5.5 provides an alternative proof of Theorem 6.5 of [V].

\vskip 20pt
\noindent J.W. van de Leur

\noindent Faculty of Applied Mathematics

\noindent University of Twente

\noindent P.O. Box 217

\noindent 7500 AE Enschede

\noindent The Netherlands

\enddocument